\documentclass[a4paper,10pt]{article}

\usepackage{color,a4wide,authblk}

\usepackage{amssymb,amsmath,amsthm}
\usepackage{graphicx}
\usepackage{textcomp}
\usepackage{amstext}
\usepackage{amsfonts}
\usepackage{amsxtra}
\usepackage{hyperref}

\usepackage{txfonts}
\usepackage{url}
\usepackage{listings}
\usepackage[output-decimal-marker={.}]{siunitx}
\usepackage{xcolor}
\usepackage{comment} 

\definecolor{codegreen}{rgb}{0,0.6,0}
\definecolor{codegray}{rgb}{0.5,0.5,0.5}
\definecolor{codepurple}{rgb}{0.58,0,0.82}
\definecolor{backcolour}{rgb}{0.98,0.98,0.95}

\lstdefinestyle{mystyle}{
  backgroundcolor=\color{backcolour},   commentstyle=\color{codegreen},
  keywordstyle=\color{magenta},
  numberstyle=\tiny\color{codegray},
  stringstyle=\color{codepurple},
  basicstyle=\ttfamily\footnotesize,
  breakatwhitespace=false,         
  breaklines=true,                 
  captionpos=b,                    
  keepspaces=true,                 
  numbers=left,                    
  numbersep=5pt,                  
  showspaces=false,                
  showstringspaces=false,
  showtabs=false,                  
  tabsize=2
}

\newcommand{\ed}{\dot{\varepsilon}}

\newcommand{\tr}{\mathrm{tr} \,}
\newcommand{\vc}[1]{\boldsymbol{#1}}
\newcommand{\vt}[1]{\boldsymbol{\mathsf{#1}}}
\newcommand{\tsp}{\mathsf{T}}
\newcommand{\de}{\partial}
\newcommand{\p}{p}

\newcommand{\Rey}{\mathrm{Re}}

\title{A multi-scale method for complex flows of non-Newtonian fluids}

\author{Francesca Tedeschi$^1$, Giulio G.\ Giusteri$^1$\footnote{giulio.giusteri@math.unipd.it}, Leonid Yelash$^2$, M\'aria Luk\'a\v{c}ov\'a-Medvid'ov\'a$^2$}
\affil{$^1$Department of Mathematics ``Tullio Levi-Civita'', University of Padua, via Trieste 63 , 35131, Padova, Italy\\
$^2$Institute of Mathematics, Johannes Gutenberg University of Mainz, Staudingerweg 9, 55128 , Mainz, Germany}
\date{}

\begin{document}

\maketitle

\begin{abstract}
We introduce a new  heterogeneous multi-scale method for the simulation of flows of non-Newtonian fluids in general geometries and  present its application to paradigmatic two-dimensional flows of polymeric fluids.
Our method combines micro-scale data from non-equilibrium molecular dynamics (NEMD) with macro-scale continuum equations to achieve a data-driven prediction of complex flows.
At the continuum level, the method is model-free, since the Cauchy stress tensor is determined locally in space and time from NEMD data.
The modelling effort is thus limited to the identification of suitable interaction potentials at the micro-scale.
Compared to previous proposals, our approach takes into account the fact that the material response can depend strongly on the local flow type and we show that this is a necessary feature to correctly capture the macroscopic dynamics.
In particular, we highlight the importance of extensional rheology in simulating generic flows of polymeric fluids.
\end{abstract}


\section{Introduction}

Modelling and computational simulation of non-Newtonian fluids is a challenging problem, since these fluids  exhibit complex effects, such as shear thinning or thickening, viscoelasticity or flow-induced phase separation. A detailed analysis of the rheology of complex fluids can be obtained by particle-based simulations. Clearly, such micro-scale description is accurate but computationally very expensive and cannot be applied to engineering-scale problems. 
Consequently, new mathematical algorithms and hybrid multi-scale approaches have been proposed in recent years.

In the literature, we can find several hybrid models combining particle dynamics with a macroscopic continuum model, such as the heterogeneous multi-scale methods \cite{donev_2010,E2004, E2003, E_2005, flekkoy_2000, Yasuda_2010}, the triple-decker atomistic-mesoscopic-continuum method \cite{Fedosov_2009, Fedosov_2008}, 
the seamless multi-scale methods \cite{E2007, e_2007, ren}, the equation-free multi-scale methods \cite{Kevrekidis_2009, Kevrekidis_2003, Kevrekidis_2004}
or the internal-flow multi-scale method \cite{lockerby2, lockerby1}. In \cite{BHJ_14} software requirements and design principles are presented and illustrated for a prototype coupling between molecular dynamics and Lattice Boltzmann methods. Note that, in general, 
hybrid multi-scale approaches are successful when processes occurring on a small scale are only loosely coupled with the large-scale behavior, that is, in the presence of an effective scale separation.
In keeping with this, our method is applicable when the local flow conditions experienced by a fluid element change on a much larger timescale than the microscopic relaxation time necessary to achieve a statistically steady state of the molecular conformation and interactions.

The aim of this paper is to illustrate a new heterogeneous multi-scale method for the simulation of flows of complex fluids in generic geometries and different conditions of motion. 
We extend and generalize the idea presented in \cite{stalter2018molecular}, where a heterogeneous multi-scale method is developed for polymeric solutions subjected to simple shear flows.
Simple shear is very well investigated for it is more easily realized in experiments than other flows. It provides fundamental information that is often sufficient to characterize simple fluids. However, complex fluids show a molecular structure that is able to change with respect to different conditions of motion, geometries, but also time scales and deformation rates.
Therefore, to retrieve important rheological information about the stress response of such fluids, their behavior is studied under conditions of flow different from steady simple shear, for example extensional motions, startup flows and oscillating shear \cite{minale1997effect}. Moreover, it is also fundamental, for these types of materials, to consider flows in complex geometries that contain holes, barriers, contractions or other irregular features, because in close proximity of these structures the fluid is subject to local motions that are not equivalent to simple shear and can manifest some unexpected behavior. 

We focus our efforts on the development of a method to link the micro-scale rheological information, available from simulations in different local flow types, to macro-scale flows in complex geometries. 
The main aim of this work is to show that it is necessary to correctly take into account the local flow type, which plays an important role in modifying the stress response in non-viscometric flows of non-Newtonian fluids featuring a flow-type dependent rheology, such as polymeric fluids.
While simulations based on constitutive assumptions do take into account, by construction, the possible flow type dependence incorporated at the level of constitutive laws, multi-scale methods for fluid mechanics application has so far considered simple shear flows as the only source of information for the micro-to-macro coupling.
With our approach, we overcome this severe limitation.
The method is applied to planar flows, but it can be extended to more general three-dimensional flows.
The most challenging task for such an extension will be to implement MD simulations under arbitrary local flow conditions.

The paper is organized in the following way. Section~\ref{Sec2} discusses the essential question 
that arises in building a heterogeneous multi-scale method. Namely, 
how micro- and macroscopic models are linked together.
Section~\ref{Sec3} is devoted to the description of micro-scale simulations with an emphasis on the decomposition of the average stress tensor and the reconstruction of data-driven material functions. For the macroscopic simulation of incompressible flows we use, as described in Section~\ref{Sec4}, a numerical method based on a semi-implicit time evolution scheme and mixed $\mathsf{P}_2$-$\mathsf{P}_0$ 
finite elements for the approximation of the velocity-pressure pair under the incompressibility constraint. Results of numerical simulations obtained by our heterogeneous multi-scale method in different geometries are presented in Section~\ref{Sec5}. The paper is closed by Section~\ref{Sec6} with an outlook on future research.

\section{Link between micro-scale data and macro-scale simulations}
\label{Sec2}

At the heart of our proposal there is the understanding that the local kinematics of a generic planar flow must be identified by \emph{at least} two parameters, one measuring the speed of the deformation and the other measuring the local flow type (see~\cite[Chapter 1.4]{larson1999structure} for a discussion of this point). 
We introduce the symmetrized gradient $\vt D\equiv\tfrac{1}{2}(\nabla\vc u+\nabla\vc u^\tsp)$ of the velocity field $\vc u$ and, for the case of planar flows in which $\vc e_3$ is normal to the flow plane, we can define the kinematic parameters
\[
|\ed|\equiv\sqrt{\frac{\tr\vt D^2}{2}}\qquad\text{and}\qquad \beta_3\equiv\frac{1}{2|\ed|}\nabla\times\vc u\cdot\vc e_3.
\]

While there is ample agreement about measuring the deformation rate in generic situations by means of $\ed$, different choices can be made for the flow-type parameter, depending on the type of material response that we envision (see~\cite{giusteri2018theoretical}, Sec.~III).
Here we choose $\beta_3$ for its simplicity and we stress that, to make it a frame-indifferent parameter, we consider it as measured relative to the vanishing spin of a fixed inertial frame of reference.
In other contexts, different choices may be appropriate (such as the relative rotation rate proposed by Schunk and Scriven~\cite{schunk1990constitutive}), but the structure of the scheme that we are going to present remains the same.

The local kinematic parameters $\ed$ and $\beta_3$ are the macro-scale input of micro-scale simulations that will in turn give an ensemble-averaged stress tensor $\vc\sigma$. 
To be able to use the micro-scale information encoded in $\vc\sigma$ we must express it in a form that is independent of the coordinates, because coordinates that are chosen for computational convenience cannot always be coherent when we pass from the micro- to the macro-scale simulation.
We can achieve this independence by decomposing the stress on an orthogonal tensorial basis built upon $\vt D$ (see~\cite{giusteri2018theoretical} for details and significance of this decomposition).
For the case of planar incompressible flows, we have
\begin{equation}
\label{stressdeco}
    \vc\sigma= -\p \vt I + 2 \eta\vt D + 2 \lambda_3 \vt G_3,   
\end{equation}
where $\vt I$ is the identity tensor and
\begin{equation}
\label{g3new}
    \vt G_3 = \tfrac{1}{2} \left(\vt A\vt D- \vt D\vt A\right) \quad \text{with} \quad \vt A= \begin{bmatrix}0 & -1 \\ 1 & 0 \end{bmatrix}. 
\end{equation}
The coefficients of the linear decomposition~\eqref{stressdeco} represent a pressure, a generalized viscosity and a reorientation factor related to normal stress differences and can be retrieved from the computational data in any coordinates by computing
\begin{equation}
\label{projection}
    \p = - \frac{\vc\sigma:\vt I}{\Vert\vt I\Vert^2} = -\frac{1}{3} \tr\left(\boldsymbol{\sigma}\right), \qquad 
    \eta = \frac{1}{2} \frac{\vc\sigma:\vt D}{\Vert\vt D\Vert^2}, \qquad \lambda_3 = \frac{1}{2} \frac{\vc\sigma:\vt G_3}{\Vert\vt G_3\Vert^2}.
\end{equation}
Here the double scalar product and the tensorial norm are defined by
\[
\vt C:\vt B\equiv\tr(\vt C^\tsp\vt B)\qquad\text{and}\qquad
\Vert\vt B\Vert\equiv\sqrt{\vt B:\vt B}.
\]

For the sake of clarity, we indicate with $\tilde{\eta}$ and $\tilde{\lambda}_3$ the two material functions of the kinematic parameters $(\ed, \beta_3)$, reconstructed by sampling the two-dimensional parameter space. 
In this way, we obtain from computational rheological measurements the constitutive  expression 
\begin{equation}
\label{stressreco}
\vc\sigma = -\p \vt I + 2 \tilde{\eta}(\ed, \beta_3) \vt D + \tilde{\lambda}_3(\ed, \beta_3) \vt G_3 
\end{equation}
for the stress tensor, that can be used in performing the macro-scale continuum simulations.
There is no need to define a pressure function because it is determined, at the macro-scale, as the Lagrange multiplier of the incompressibility constraint.

\begin{figure}
    \centering
    \includegraphics[width=0.90\textwidth]{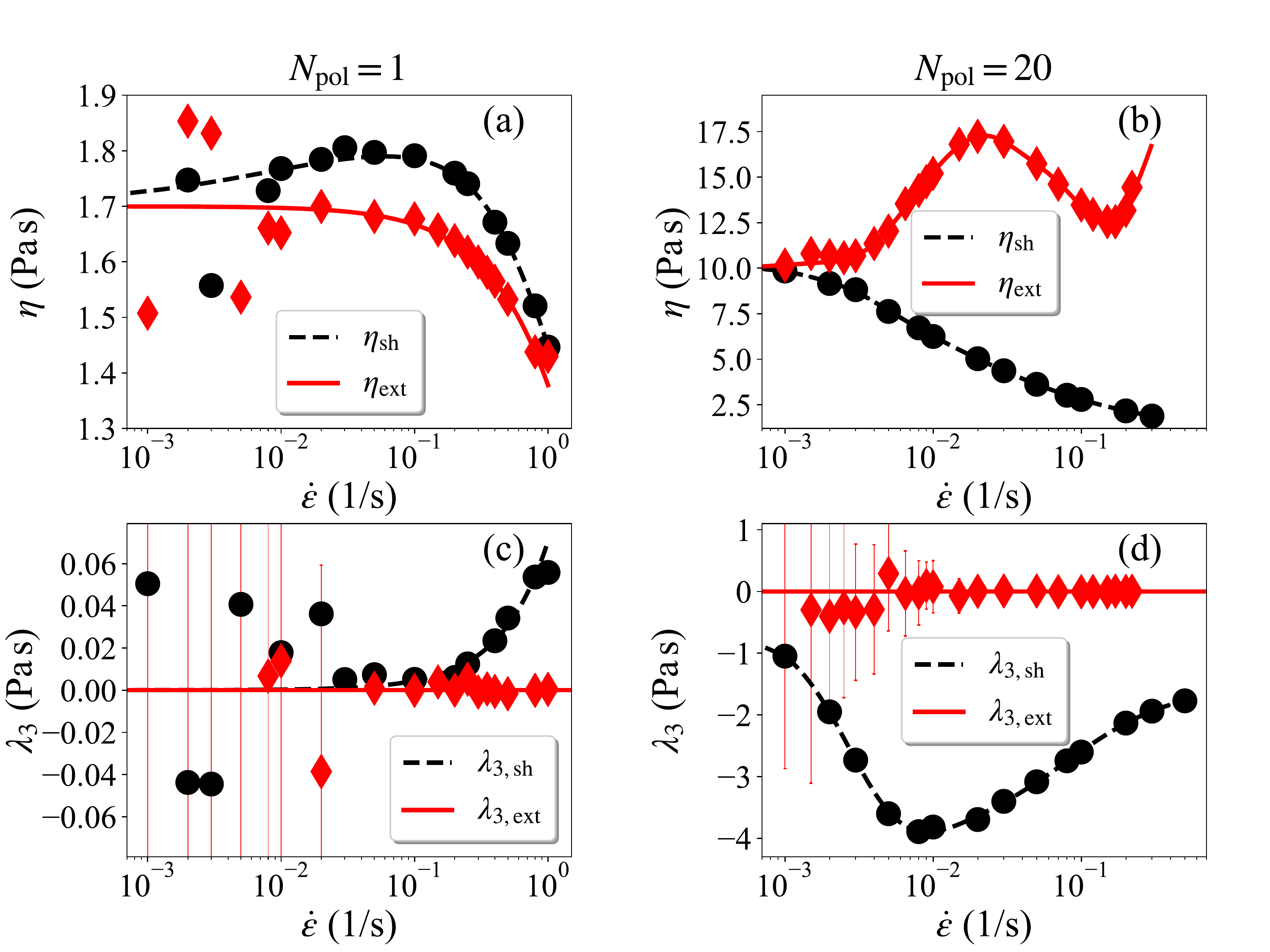}
    \caption{\emph{The monomeric fluid displays almost no rate dependence and very little flow-type dependence of the stress coefficients, while the polymeric fluid features rate dependence and a strong flow-type dependence.}
    We compare the NEMD data for the stress coefficients $\eta$ (top) and $\lambda_3$ (bottom) against $\ed$ obtained for the monomeric case with $N_\mathrm{pol}=1$ (left) and the polymeric case $N_\mathrm{pol}=20$ (right) under simple shear (circles and black dashed curves) and planar extension (diamonds and red solid curves). In the latter case, $\lambda_3$ fluctuates around zero, that is the expected value based on symmetry considerations. The presence of a strong flow-type dependence for the polymeric fluid is a key feature, originated by the different conformations of the long molecules. Fitting curves are obtained as described in the text.}  
\label{fig:eta}
\end{figure}

\section{Micro-scale simulation}
\label{Sec3}

The NEMD simulation at the micro-scale is performed using LAMMPS~\cite[lammps.sandia.gov]{plimpton1995fast}.
We consider both monomeric and polymeric aggregates in a three-dimensional computational domain with periodic boundary conditions and undergoing two different planar flows: simple shear and extensional flow.
The average velocity field is imposed by suitably deforming the computational box and the canonical ensemble statistics of the velocity fluctuations is achieved via the classical Nos\'e--Hoover thermostat~\cite{frenkel2001understanding}.
For the case of simple shear we use the LAMMPS command \verb|fix deform| in conjunction with the SLLOD algorithm~\cite{daivis2006simple}, while the extensional flow is treated by means of the UEF package~\cite{nicholson2016molecular}.
The latter implements boundary conditions developed by Dobson~\cite{dobson2014periodic} and  Hunt~\cite{hunt2016periodic}, that are an extension of the boundary conditions of Kraynik and Reinelt~\cite{kraynik1992extensional}. 
We will discuss results for monomers, $N_\mathrm{pol}=1$, and polymers, $N_\mathrm{pol}=20$, where $N_\mathrm{pol}$ is the number of monomers per molecule.

\subsection{Interaction potentials}

We model polymers as bead-spring systems with two types of interaction potential. The first one is active between each pair of monomers and is the Weeks-Chandler-Andersen (WCA) potential, the repulsive part of a Lennard-Jones potential with minimum energy $-\epsilon$ when  the centers of the beads are at distance $r=\sqrt[6]{2}\sigma$:
\[
\varphi_\mathrm{WCA}(r)\equiv\left\{\begin{aligned}
&4\epsilon\bigg[\bigg(\frac{\sigma}{r}\bigg)^{12}-\bigg(\frac{\sigma}{r}\bigg)^{6}\bigg]+\epsilon& &\text{if }r<\sqrt[6]{2}\sigma\\
&0 & &\text{otherwise.}
\end{aligned}\right.
\]
The second one, only active between successive monomers in a polymeric chain, is the Finitely Extensible Nonlinear Elastic (FENE) potential
\[
\varphi_\mathrm{FENE}(r) \equiv -0.5 K R_0^2 \ln{\left[1- \left( \frac{r}{R_0}\right)^2\right]},
\]
with $K$ the elastic constant and $R_0$ the maximum extent of the bonds.

\begin{figure}[h]
    \centering
    \includegraphics[width=0.90\textwidth]{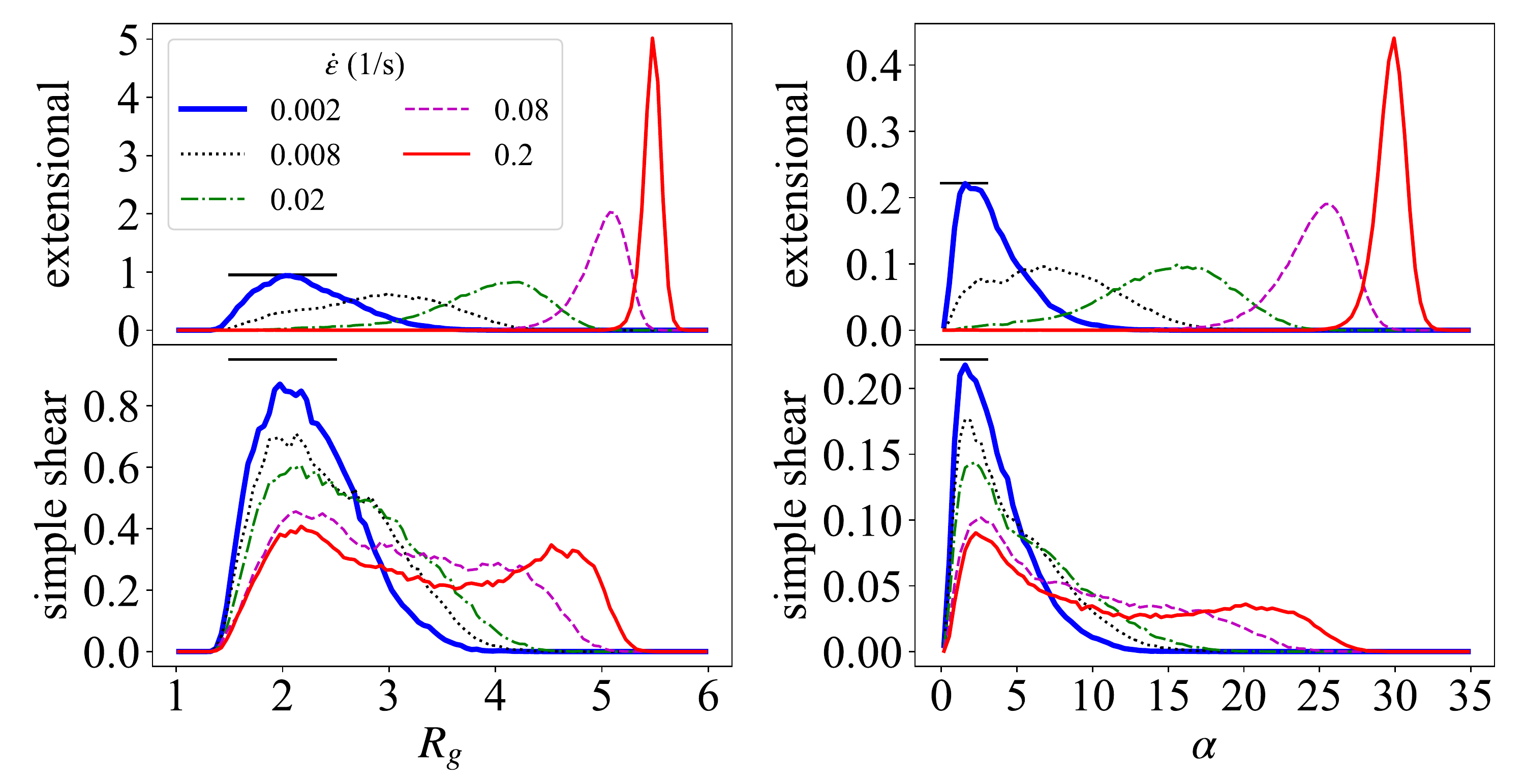}
    \caption{\emph{The presence of a strong flow-type dependence for the polymeric fluid is a key feature, originated by the different conformations of the long molecules.} The probability distribution of radius $R_g$ of gyration (divided by the effective monomer radius) and asphericity $\alpha$ for the polymeric chains under simple shear and extensional flow is reported for different values of $\ed$. The black bars indicate equal values for an easier comparison between the results under simple shear and planar extension. The curves are generated considering conformational data of five chains over 20,000 time steps of steady-state NEMD.}  
\label{fig:gyr}
\end{figure}

\subsection{Average stress coefficients}

We considered values of the deformation rate $\ed$ ranging from 0.001 s$^{-1}$ to 1 s$^{-1}$.
For each value of $\ed$ we performed 21 simulations with different initial configurations of the particles.
At each time, the average stress given as output by LAMMPS is projected onto the basis tensors through the formulae \eqref{projection} and these projections are averaged over time. Finally, a mean over all initial configurations is taken.
Figure~\ref{fig:eta} presents the data obtained for the generalized viscosity $\eta$ and the reorientation factor $\lambda_3$. 
By fitting those data with Carreau and power functions we obtain, for the monomeric case with $N_\mathrm{pol}=1$, the curves
\begin{align*}
&\lambda_\mathrm{3,ext}(\ed)=0,
&\eta_\mathrm{ext}(\ed)=1.7/\big[1+(0.108\,\ed)^{1.05}\big]^{2.276},\\
&\lambda_\mathrm{3,sh}(\ed)=0.0706\,\ed^{1.2},
&\eta_\mathrm{sh}(\ed)=\frac{1.806}{\big[1+(1.6\,\ed)^{1.65}\big]^{0.191}}+\frac{1.7-1.806}{\big[1+(0.3\,\ed)^{0.53}\big]^{22}}.
\end{align*}
For the polymeric case with $N_\mathrm{pol}=20$, the Gaussian Process Regression method (see Appendix~\ref{appendice}) was used to obtain the fitting curves.

By comparing the results obtained under simple shear and planar extension for $N_\mathrm{pol}=1$ and $N_\mathrm{pol}=20$, we find that the former fluid displays almost no rate dependence and very little flow-type dependence of the stress coefficients, while the latter features rate dependence and a strong flow-type dependence (Figure \ref{fig:eta}).
The statistics of interaction between monomers in the two types of flow does not differ significantly, leading to a quasi-Newtonian rheology.
On the contrary, the fact that polymer chains are kept in an elongated state by extensional flows, while they tend to tumble and keep a spherical shape in simple shear, generates a richer phenomenology.
In fact, the shear-thinning trend displayed in simple shear is qualitatively similar to that of monomeric fluids, while we find an opposite trend---a shear-thickening behavior---in extensional flows at small deformation rates, followed by a non-monotonic behavior of the viscosity.
The elastic effects generated by the bonds are crucial in this case and give rise, in simple shear, to the normal stress differences associated with the coefficient $\lambda_3$, also featuring a non-monotonic curve. 

The direct connection between the statistics of chain conformation in the polymeric case and the observed rheology can be evinced from the probability distribution, presented in Figure~\ref{fig:gyr}, of the normalized radius of gyration $R_g$ and asphericity $\alpha$ defined as follows. Let the components of the $3\times3$ gyration tensor be
\[
\vt R_{\mu\nu}\equiv \frac{1}{M}\sum_{i=1}^{20}m_i(\vc r_i-\vc r_\mathrm{cm})_\mu(\vc r_i-\vc r_\mathrm{cm})_\nu,
\]
with $m_i$ the mass of the $i$-th monomer, $\vc r_i$ its position, $M=\sum_im_i$, and $r_\mathrm{cm}$ the center of mass of the polymeric chain. 
With $R_0$ the effective monomer radius, we set
\[
R^2_g\equiv \frac{1}{R_0^2M}\sum_{i}m_i|\vc r_i-\vc r_\mathrm{cm}|^2=\frac{\tr\vt R}{R_0^2}.
\]
Moreover, the asphericity is given by
\[
\alpha\equiv \frac{1}{R_0^2}\bigg[a_3-\frac{1}{2}(a_1+a_2)\bigg],
\]
where $a_1\leq a_2\leq a_3$ are the eigenvalues of the gyration tensor $\vt R$.

At low deformation rate, the distributions obtained in simple shear and extensional flows are almost identical (blue curves in Figure~\ref{fig:gyr}) because the chains are equally and mildly stretched by the imposed flow. As $\ed$ grows larger, the extensional flow distributions feature a single peak that is progressively shifted rightward, toward more stretched configurations. On the other hand, in simple shear we notice a broadening of the distribution eventually leading to a doubly peaked shape (red curves in Figure~\ref{fig:gyr}). This indicates that chains spend roughly half of the time in a stretched configuration while frequently going back to a coiled configuration.
In this very different microscopic behavior we find the origin of the strong flow-type dependence of the viscosity at larger $\ed$.

\subsection{Data-driven material functions}

Ideally, we would like to perform micro-scale simulations with many values of $\ed$ and $\beta_3$, since flows in generic geometries can easily feature variations of the flow type with $\beta_3$ ranging from minus to plus infinity and values of $\ed$ that cover many orders of magnitude.
The implementation of simulation algorithms for mixed flows different from simple shear and planar extension is a nontrivial task that will be addressed in future works, but we still need, for our macro-scale simulations, a suitable definition of material functions that covers the whole range of kinematic parameters.
We thus apply simple extrapolation strategies to build material functions out of the available computational data. 

Specifically, we extend the fitted curves $\eta_\mathrm{sh}(\ed)$, $\eta_\mathrm{ext}(\ed)$, $\lambda_\mathrm{3,sh}(\ed)$ and $\lambda_\mathrm{3,ext}(\ed)$ out of their natural domains by taking constant values.
Then we set
\begin{equation}\label{eq:eta}
\tilde{\eta} (\ed, \beta_3)\equiv \left\{
\begin{aligned}
& (1-|\beta_3|) \eta_\mathrm{ext}(\ed) + |\beta_3| \eta_\mathrm{sh}(\ed),  \quad &\beta_3 \in [-1,1]\\
&\eta_\mathrm{sh}(\ed),  \quad &|\beta_3|>1\end{aligned}\right.
\end{equation}
that is an even function of $\beta_3$ with $\eta_\mathrm{ext}(\ed)= \tilde{\eta}(\ed, 0)$ and  $\eta_\mathrm{sh}(\ed)= \tilde{\eta}(\ed, 1)$, and
\begin{equation}\label{eq:lambda}
\tilde{\lambda}_3(\ed, \beta_3)\equiv \left\{
\begin{aligned}& \beta_3 \lambda_\mathrm{3 ,sh}(\ed),  \quad &\beta_3 \in [-1,1]\\
&\pm\lambda_\mathrm{3,sh}(\ed),  \quad &\beta_3 \gtrless \pm 1 \end{aligned}\right.
\end{equation}
that is an odd function of $\beta_3$ and such that $\lambda_\mathrm{3, ext}(\ed)= \tilde{\lambda}_3(\ed, 0) \approx 0$ and $\lambda_\mathrm{3 ,sh}(\ed)= \tilde{\lambda}_3(\ed, 1)$.
These operations in combination with \eqref{stressreco} allow us to build a complete set of data for the stress tensor as a function of $\ed$ and $\beta_3$.

\section{Macro-scale simulation}
\label{Sec4}

We consider the flow of an incompressible fluid with constant density $\rho$ and velocity vector $\vc{u}$. 
From the macroscopic point of view the continuum description leads to the conservation of mass and the balance of linear momentum. In the incompressible case, they translate into the following partial differential equations on the domain $\Omega$:
\begin{align}
       \nabla \cdot \vc{u}&= 0,    \label{mass} 
       \\
    \rho \left(\frac{\de \vc{u}}{\de t} + \left(\vc{u}\cdot \nabla \right)\vc{u}\right)&= \nabla \cdot \vc\sigma(\vc{u}, \p). \label{motion}
 \end{align}
Here $\vt{\vc\sigma}$ denotes the Cauchy stress tensor and $p$ the pressure field. Equation \eqref{mass} expresses the incompressibility constraint.
We partition the boundary of the domain as disjoint union of inlet $\Gamma_\mathrm{in}$, outlet $\Gamma_\mathrm{out}$, and solid walls $\Gamma_\mathrm{w}$, so that $\de \Omega =\Gamma_\mathrm{in}\cup \Gamma_\mathrm{out}\cup \Gamma_\mathrm{w}$.
As usual, $\vc n$ is the outward unit normal to $\de \Omega$.

A weak formulation of the problem is retrieved by multiplying  equation \eqref{motion} by a test function $\vc v \in H^1_{\Gamma_\mathrm w}(\Omega)=\big\{\vc f \in H^1(\Omega) : \vc f\vert_{\Gamma_\mathrm{w}} = \vc 0\big\}$ and integrating over $\Omega$. By applying Green's formula,
we obtain the integral equation
\begin{equation}\label{eq:weak1}
    \int_{\Omega} \rho \frac{\de\vc u}{\de t} \cdot {\vc v} +  \int_{\Omega}\rho  ({\vc u} \cdot \nabla){\vc u}\cdot {\vc v}  = \int_{\de \Omega} 
    \boldsymbol{\sigma}\vc n\cdot{\vc v}
    - \int_{\Omega} {\boldsymbol{\sigma}}:\nabla{\vc v} .
\end{equation}
Moreover, \eqref{mass} can be multiplied by a scalar test function $q \in L^2_0(\Omega)=\big\{g\in L^2(\Omega):\int_\Omega g=0\big\}$ and integrated over $\Omega$ to give
\begin{equation}\label{eq:weak2}
\int_{\Omega} q \nabla\cdot{\vc u} = 0.
\end{equation}

We decompose the Cauchy stress tensor in spherical and deviatoric parts by introducing the traceless extra stress $\boldsymbol{\tau}$ such that $\boldsymbol{\sigma}=-(p+\bar{p}){\vt I} + \boldsymbol{\tau}$, where $\bar{p}\in L^2_0(\Omega)$ is a given pressure field used to impose a chosen pressure gradient from inlet to outlet.
On top of the Dirichlet boundary conditions for $\vc u$ on $\Gamma_\mathrm{w}$ we assume a form of periodicity for $\vc u$ and $p$, requiring that they take the same values on $\Gamma_\mathrm{in}$ and $\Gamma_\mathrm{out}$. 
Thanks to the independence of the test functions $\vc v$ and $q$, we can take the sum of \eqref{eq:weak1} and \eqref{eq:weak2} and substitute the expression for $\boldsymbol{\sigma}$ to obtain
the complete weak formulation of the problem: 

\emph{Find} $\vc u \in L^2(\mathbb{R}^+;H^1_{\Gamma_\mathrm w}(\Omega))$ \emph{and} $p \in L^2(\mathbb{R}^+;L_0^2(\Omega)),$ \emph{such that for any suitable test functions $\vc v$ and $q$ we have}
\begin{equation}\label{eq:weak3}
    \int_{\Omega} \rho \frac{\de\vc u}{\de t} \cdot {\vc v} +  \int_{\Omega}\rho  ({\vc u} \cdot \nabla){\vc u}\cdot {\vc v} + \int_{\Omega} {\boldsymbol{\tau}}:\nabla{\vc v} + \int_{\Omega} q \nabla\cdot{\vc u} - \int_{\Omega} (p+\bar{p}) \nabla\cdot{\vc v} = \int_{\Gamma_\mathrm{in}\cup\Gamma_\mathrm{out}} 
    \bar{p}\vc n\cdot{\vc v}
     .
\end{equation}
Equation~\eqref{eq:weak3} must be completed with suitable initial conditions for the velocity field, that we will assume to vanish identically at time $t=0$.
To exploit the reflection symmetry of some domains, we will consider a fictitious boundary $\Gamma_\mathrm{c}$ corresponding to the symmetry axis. On this boundary we will assume that the normal component of the velocity field and the tangential component of the traction $\boldsymbol{\tau}\vc n$ vanish.

For the time integration of \eqref{eq:weak3} we apply  a semi-implicit Euler scheme. In particular, the nonlinear convective term  is approximated explicitly, the rest is approximated implicitly. This approximation is suitable to simulate flows at sufficiently low Reynolds number. Consequently, we have
\begin{equation}\label{eq:weak_dis}
    \int_{\Omega}\rho \frac{{\vc u}^{n+1}}{\Delta t} \cdot {\vc v} + a({\vc u}^{n+1}, {\vc v}) + b({\vc v}, p^{n+1}+\bar{p}) - b({\vc u}^{n+1}, q) = \int_{\Omega}\rho \frac{{\vc u}^{n}}{\Delta t} \cdot {\vc v} - \int_{\Omega}\rho  ({\vc u^n} \cdot \nabla){\vc u^n}\cdot {\vc v}  + \int_{\Gamma_\mathrm{in}\cup\Gamma_\mathrm{out}} 
    \bar{p}\vc n\cdot{\vc v}
\end{equation}
with bilinear forms $b(\vc{u}, q):= -\int_{\Omega} q \nabla\cdot{\vc u}$ and
$a(\vc{u}, \vc{v}):= \int_{\Omega}{\boldsymbol{\tau}(\vc u)}:\nabla{\vc v}.$ Note that the latter is already linearized since the (nonlinear) material functions in \eqref{stressreco} are computed using  $\vc{u}^n.$

For each time step, Equation~\eqref{eq:weak_dis} is discretized in space and solved, in a standard way, by means of mixed finite elements $\mathsf{P}_2$-$\mathsf{P}_0$  for the approximation of the velocity-pressure pair, which are known to satisfy the Ladyzhenskaya--Babu\v{s}ka--Brezzi inf-sup condition for a stable solution of the associated saddle-point problem~\cite{brezzi2012mixed}.
The numerical method is implemented within the Python library FEniCS~\cite{logg2010dolfin} and some details pertaining the micro-to-macro coupling through the dynamic reconstruction of the extra stress are given in Appendix~\ref{app:code}.

\section{Numerical results for different geometries}
\label{Sec5}

The aim of this section is to demonstrate the robustness and effectiveness of our heterogeneous multi-scale method  by presenting the results of simulations in three different planar geometries: flows in a straight channel, through a 4:1 contraction, and past a deep hole. These are paradigmatic geometries frequently used to test the non-Newtonian behavior of fluid models~\cite{crochet2012numerical}.

\begin{figure}[t]
    \centering
    \includegraphics[width=0.9\textwidth]{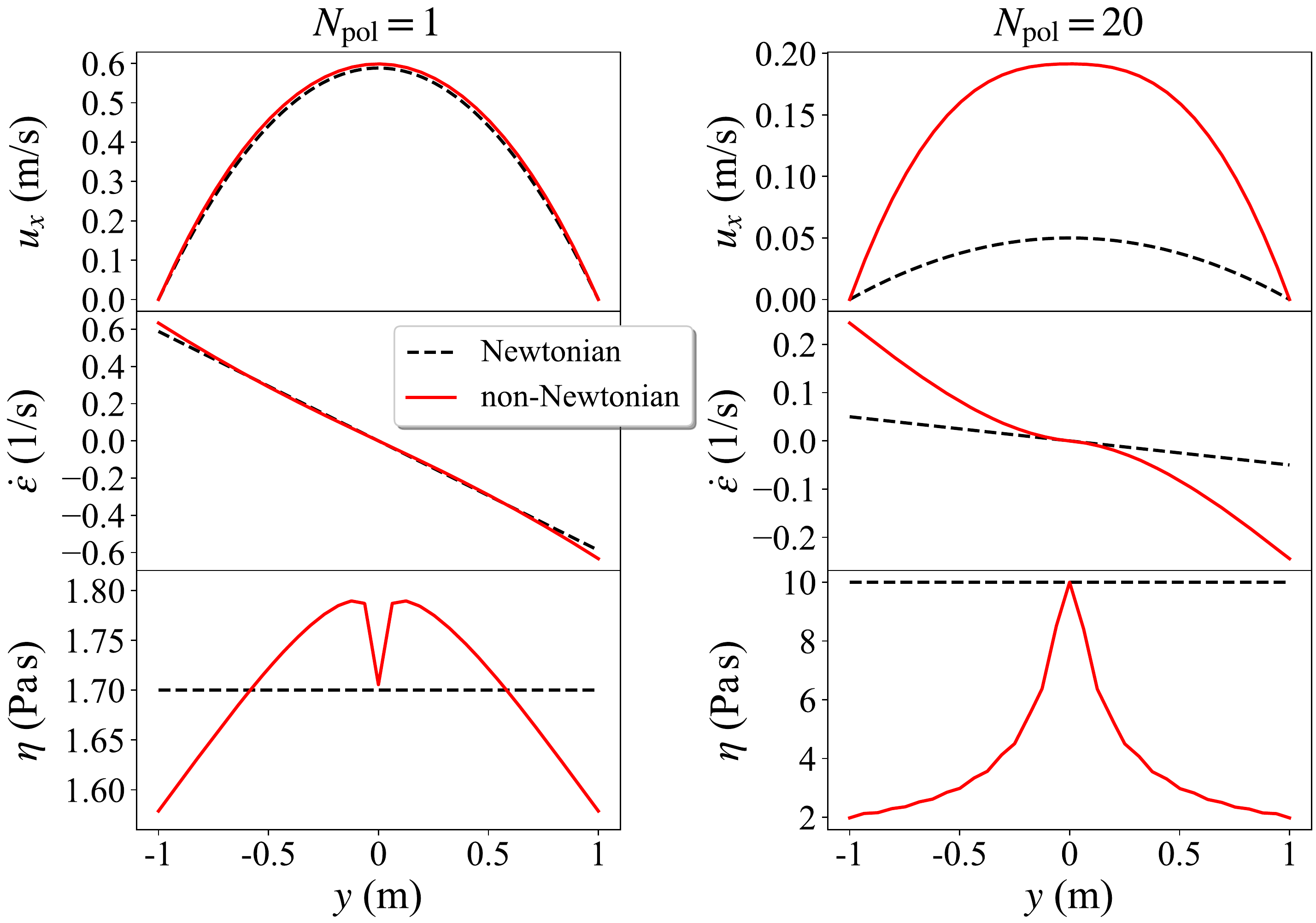}
    \caption{%
    {\emph{Comparison of the horizontal velocity profile $u_x$, local deformation rate $\ed$, and viscosity $\eta$ for the flow in a straight channel computed by our multi-scale method with a reference Newtonian solution.}  
    In the case $N_\mathrm{pol}=1$, the non-Newtonian solution is extremely close to the Newtonian one, as expected from the mild rate dependence of the viscosity and the small values of $\lambda_3$. In the case $N_\mathrm{pol}=20$, we observe a much larger deviation with a shear-thinning character expected from the rate dependence of the viscosity.}}
    \label{fig:channel}
\end{figure} 
\subsection{Channel flow}

The multi-scale method is firstly tested on the classical case of a planar channel flow. The total length of the channel is $L= \SI{5}{\meter}$ along the $x$-axis and the system width is $W=\SI{2}{\meter}$ along the $y$-axis.
Exploiting the reflection symmetry of the problem, we discretized only the upper half of the channel on a mesh with $100$ triangular elements along the $x$-axis and $20$ elements along the $y$-axis, for a total of $4000$ triangles in the entire computational domain.
On the fictitious boundary that corresponds to the center of the channel we impose a vanishing vertical velocity and a vanishing normal traction to respect the reflection symmetry of the problem.
We consider a fluid with unit density and drive the flow by imposing a constant horizontal pressure gradient $C=\SI{1}{\pascal/\meter}$. 
As for the velocity field $\vc u$, we assume periodic boundary conditions so that the velocity at inlet and outlet is the same.
The time step is $\Delta t =\SI{1d-3}{\second}$.
We let the system evolve up to the time $T=\SI{2}{\second}$ to reach a steady state.

The flow type is everywhere equivalent to that of a simple shear. Indeed, this is a viscometric flow with $\beta_3=1$ in the top half of the domain and $\beta_3=-1$ in the other half. A discontinuity would appear at the midline of the channel consistent with the fact that $\beta_3$ is not defined for $\ed=0$. 
The homogeneity of the flow type makes the flow-type dependence of the stress tensor irrelevant.
To assess the importance of non-Newtonian effects, we compare the solution obtained from the multi-scale approach with the analytical solution of a reference Newtonian model, represented by the classical stress-strain relation
\begin{equation}
\vc\sigma= -\p\vt{I}+2 \bar{\eta} \vt{D}
\end{equation} 
with $\bar{\eta}$ constant. We take $\bar{\eta}=\SI{1.7}{\pascal.\second}$ for $N_\mathrm{pol}=1$ and $\bar{\eta}= \SI{10}{\pascal.\second}$ for $N_\mathrm{pol}=20$, corresponding to the zero-rate limit of the interpolated shear viscosity from our NEMD simulations.

The comparison of velocity, deformation rate, and viscosity profiles is reported in Figure~\ref{fig:channel}. In the case $N_\mathrm{pol}=1$ the non-Newtonian solution is extremely close to the Newtonian one, as expected from the mild rate dependence of the viscosity and the small values of $\lambda_3$. In the case $N_\mathrm{pol}=20$, we observe a much larger deviation that is small only around the center of the channel, where the shear rate vanishes and the viscosity approaches that of the reference Newtonian model.

\begin{figure}[t] 
\centering
\includegraphics[width=0.91\textwidth]{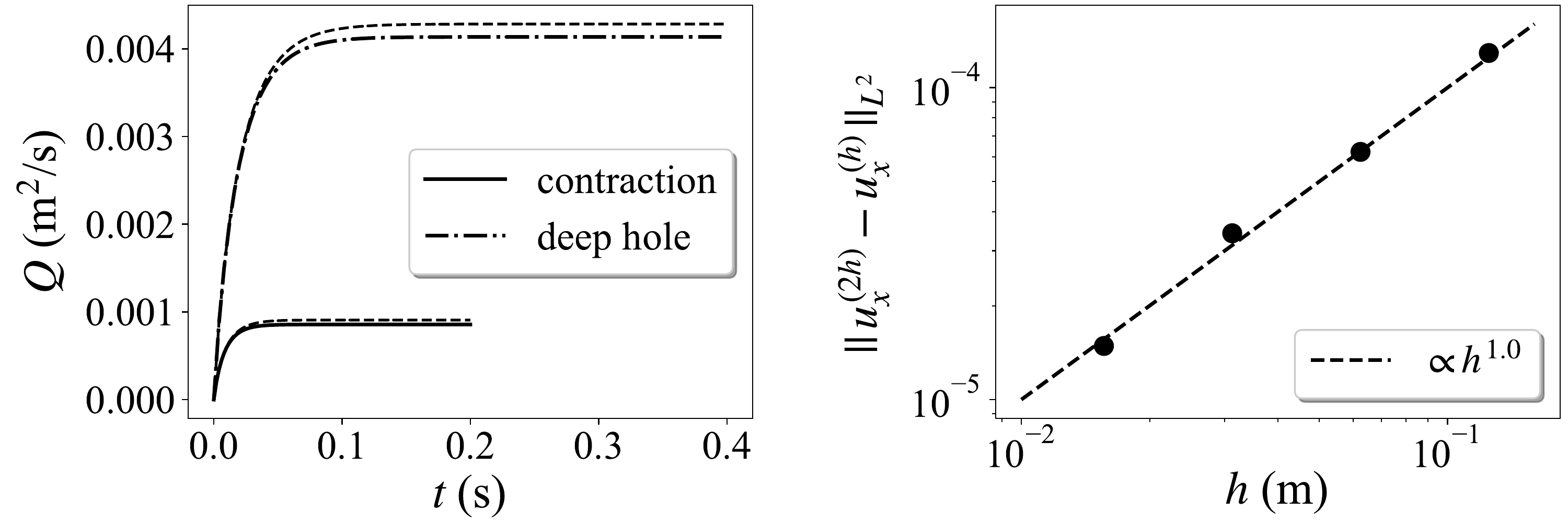}
\caption{\emph{Time convergence of the numerical solution to the steady flow (left) and mesh convergence of the method (right).}
The attainment of steady flow conditions is confirmed by the evolution of the flow rate $Q$ through vertical sections of the domain for the contraction and deep hole geometries (left, see legend) the dashed lines show the flow rate produced by the modified non-Newtonian model and highlight a measurable difference with the full model.
The incremental correction obtained by successive mesh refinements is measured by the $L^2$ norm of the difference of the corresponding velocity fields and features a linear scaling with the mesh size parameter $h$, thus showing convergence of the method in a situation in which all the non-Newtonian contributions are activated.}
\label{fig:timeandmesh}
\end{figure}

Due to the non-uniformity of the viscosity in the development of the flow and in its steady state, the characterization of the flow by means of dimensionless quantities such as the Reynolds number is not straightforward. 
Nevertheless, we can identify the range of local parameter values as follows.
We take the channel width $W$ as reference length and $P=CW$ as reference pressure drop. The local Reynolds number is then given by
\[
\Rey=\frac{W\sqrt{\rho CW}}{\eta}.
\]
Another important dimensionless quantity is the reorientation angle $\varphi$, such that
\[
\tan\varphi=\frac{\lambda_3}{\eta+\sqrt{\eta^2+\lambda_3^2}},
\]
which measures the rotation of the stress eigenvectors with respect to the eigenvectors of $\vt D$. This geometric information is precisely the meaning of the first normal stress difference in simple shear flows, extended to generic local flows by the definition of $\lambda_3$ and $\eta$ \cite{giusteri2018theoretical}.

In the monomeric case, we have $\eta_\mathrm{min}=\SI{1.6}{\pascal.\second}$ and $\eta_\mathrm{max}=\SI{1.8}{\pascal.\second}$ so that $\Rey\in[1.57,1.77]$, while $\varphi$ is negligible.
In the polymeric case, we have $\eta_\mathrm{min}=\SI{2}{\pascal.\second}$ and $\eta_\mathrm{max}=\SI{10}{\pascal.\second}$ so that $\Rey\in[0.28,1.41]$.
We can then consider our examples to be at low Reynolds number.
Since $\ed_\mathrm{max}=\SI{0.24}{\second^{-1}}$, we have $\varphi\in[-22,22]$ degrees, that includes quite significant values.

\subsection{Flow through a contraction}

\begin{figure}[t] 
\centering
\includegraphics[width=0.9\textwidth]{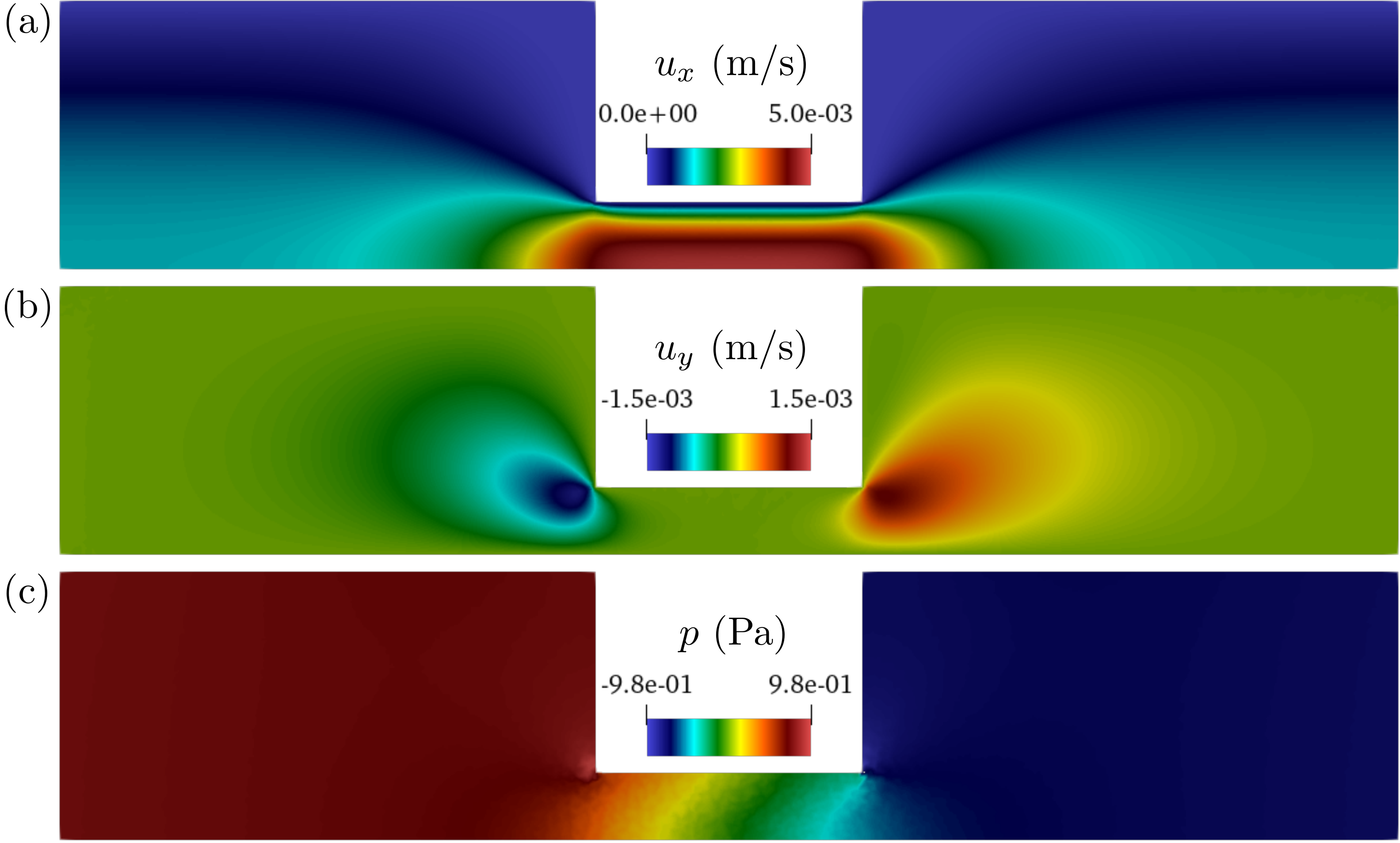}
\caption{\emph{The geometry of the domain induces a flow that features a broad spectrum of local flow types.}
(a,b) The two components of the steady-state velocity field, $u_x$ and $u_y$, in the flow through a contraction show that we are in a stable laminar regime.
(c) The pressure field features oblique isolines due to the presence of normal stress differences.}
\label{fig:streamlines_contraction}
\end{figure}

As a first example of flow in a complex geometry we consider the 4:1 contraction.
This flow is characterized by the presence of different local flow types:
simple shear in regions that are sufficiently far from the ends of the contraction, extensional and mixed motion in proximity of entrance and exit of the contraction, and 
mixed rotational flow near the corners of the domain.
The spatial non-uniformity of the flow type must be taken into account  to obtain a precise macro-scale description of the fluid motion.
To show this, we compare the prediction of the full non-Newtonian model with that of a \emph{modified} non-Newtonian model, wherein the flow-type dependence is artificially suppressed.
Specifically, $\beta_3$ is replaced by the function $\mathrm{sign}(\beta_3)$. 
In this way, only the micro-scale data obtained in simple shear (i.e., $\beta_3=\pm 1$) are used, but the results cannot properly reflect the fluid behavior.

\begin{figure}[t]
\centering
\includegraphics[width=0.9\textwidth]{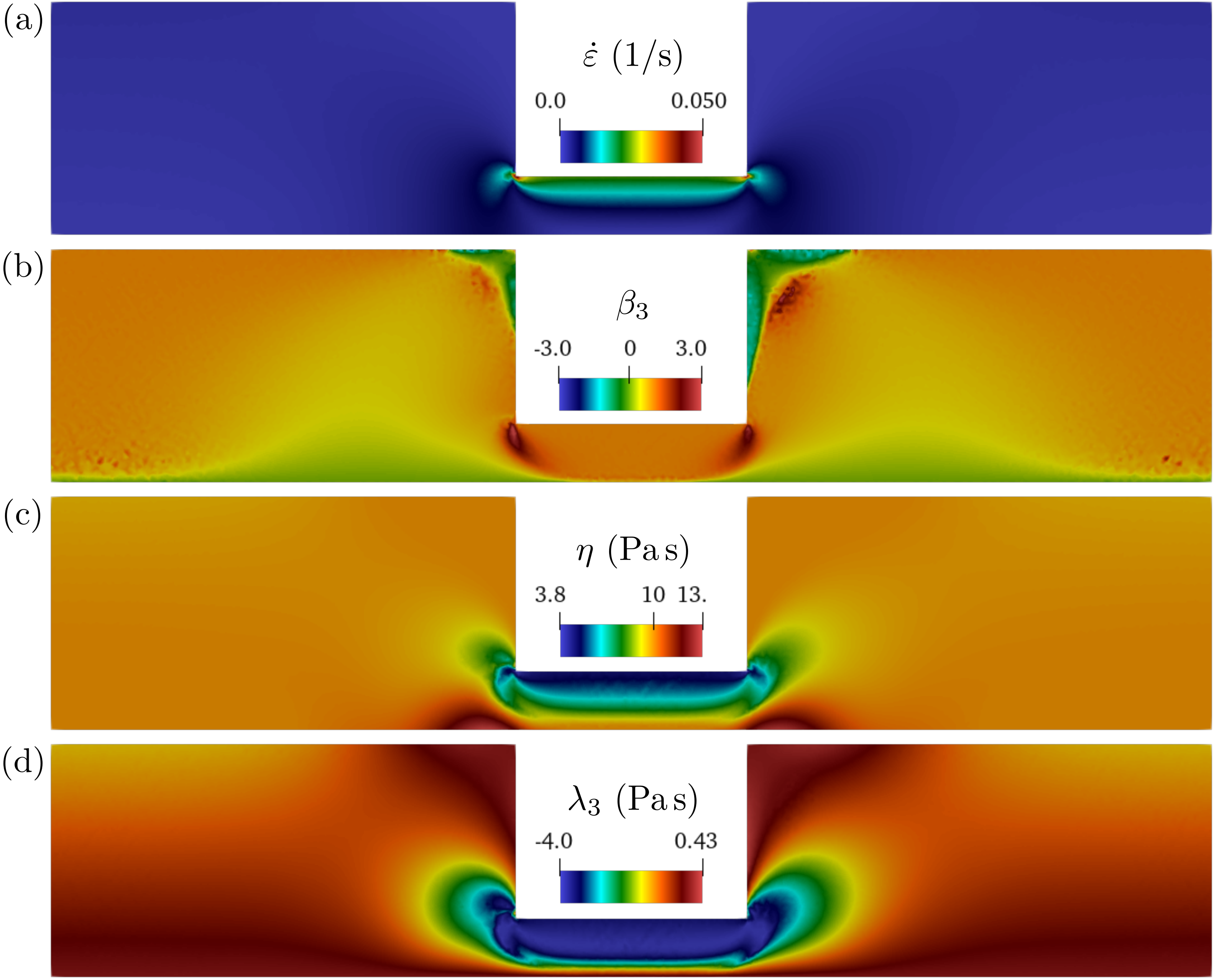}
\caption{\emph{The non-uniform flow type in complex flows can affect significantly the rheological response.} We measured (a) the deformation rate $\ed$ and (b) the local flow type $\beta_3$ in the flow through a contraction of a polymeric fluid ($N_\mathrm{pol}=20$) and computed the local values of the material parameters (c) $\eta$ and (d) $\lambda_3$. In this case, they are strongly affected by changes in the flow type on top of the variations due to the deformation rate.}
\label{fig:fields_contraction}
\end{figure}

We discuss results obtained with a channel of length $L= \SI{5}{\meter}$ and maximum width $W= \SI{2}{\meter}$, only in the polymeric case $N_\mathrm{pol}=20$, with time step $\Delta t= 10^{-3}$, letting the system evolve to reach a steady state for $T=\SI{0.2}{\second}$. We can appreciate the time convergence to the steady solution by looking at the evolution of the flow rate $Q$ through vertical sections of the domain, reported in Figure~\ref{fig:timeandmesh}, left panel.
The discretization employs an unstructured triangular mesh with average diameter of the triangles $h\approx\SI{0.03}{\meter}$. The mesh convergence of the multi-scale method is assessed by considering the $L^2$ norm of the difference between solutions obtained with mesh parameter $h$ and $2h$. 
From the data reported in Figure~\ref{fig:timeandmesh}, right panel, we see that the incremental correction scales linearly in $h$, thus showing convergence of our simulation.

The pressure gradient that drives the flow is $C= \SI{0.3}{\pascal/\meter}$.
The velocity field $\vc u$ is assumed periodic so that it is matched at inlet and outlet and, to exploit symmetry and compute the numerical solution only in the upper half of the domain, we use the same conditions as in the straight channel simulation, namely, vanishing vertical velocity and vertical traction at the center of the channel.
The flow is laminar with the characteristic increase of horizontal velocity $u_x$ in correspondence of the contraction. The pressure gradient is concentrated along the narrower portion of the domain and features oblique isolines due to the presence of normal stress differences in that region (Figure~\ref{fig:streamlines_contraction}).
The non-uniformity of the deformation rate $\ed$ and of the flow-type parameter $\beta_3$ has a strong influence on the local values of $\eta$, that range from $\SI{3.8}{\pascal.\second}$ to $\SI{13}{\pascal.\second}$ in the contracting region, and $\lambda_3$, that range from $\SI{-4}{\pascal.\second}$ to $\SI{4}{\pascal.\second}$, considering also the lower half of the domain (see Figure~\ref{fig:fields_contraction}).
From these values and taking now $W/4$ as reference length, we can estimate $\Rey\in[0.015,0.05]$ and $\varphi\in[-23.5,23.5]$ degrees.

\begin{figure}[htb]
\centering
\includegraphics[width=0.9\textwidth]{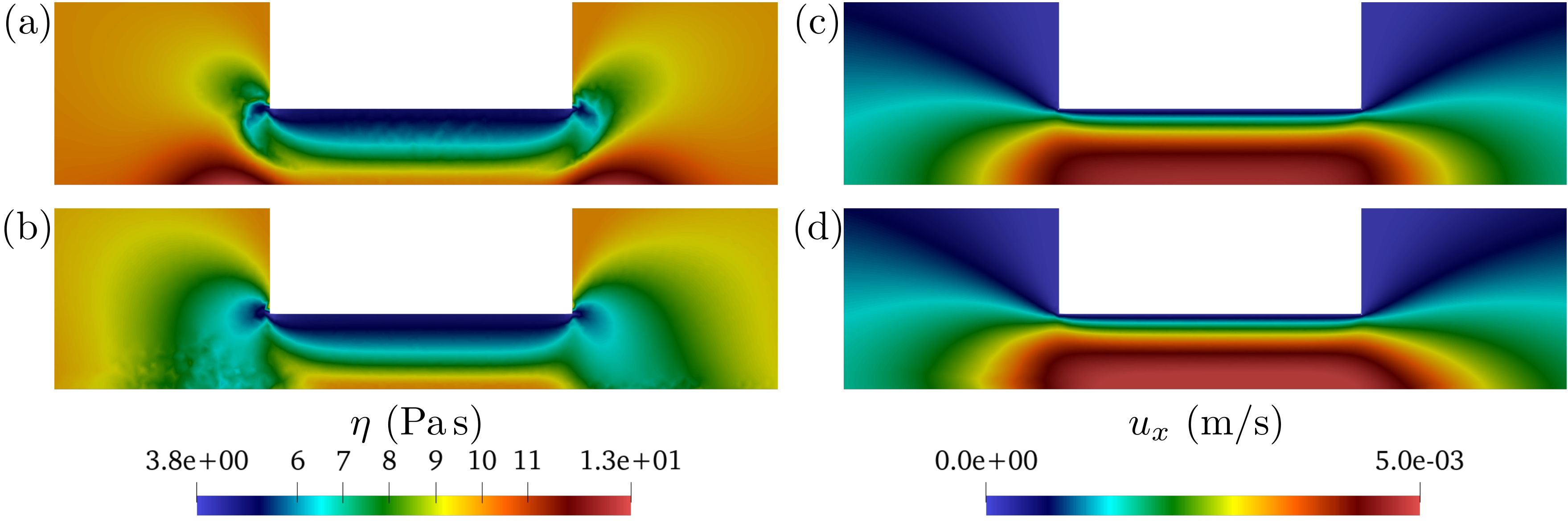}
\caption{\emph{The dependence of the stress on the local flow type must be taken into account to correctly predict the fluid behavior.}
We compare the steady-state viscosity $\eta$ and horizontal velocity $u_x$ nearby the contraction, computed with (a,c) the full non-Newtonian model and (b,d) the modified one.
The viscosity in the modified model presents a very different profile in the extensional and mixed flow regions, nearby entrance and exit of the restriction.
This influences the velocity field, which features a clear asymmetry due to the overemphasized role of normal stress differences in the modified model.}
\label{fig:sections}
\end{figure}

In this complex flow, the flow type measured by the parameter $\beta_3$ varies significantly through the domain (Figure \ref{fig:fields_contraction}b), ranging from pure extension ($\beta_3=0$) to simple shear ($|\beta_3|=1$) to more rotational flows ($|\beta_3|>1$).
Regions of mixed and extensional motion cover a major part of the domain, going well within the contraction. 
Since the viscosity in the full non-Newtonian model is a function of $\ed$ and $\beta_3$, the effect of the flow type is clearly dominant in determining its value.
The importance of correctly embedding the flow-type dependence in the multi-scale model can be highlighted by comparing the prediction of the full non-Newtonian model with that obtained from the modified non-Newtonian model (Figure \ref{fig:sections}).
The viscosity in the modified model presents a very different profile in the extensional and mixed flow regions, where it happens to decrease instead of increasing, as predicted by the correct method and in line with the extensional rheology of the polymeric case.
This has an immediate effect on the velocity profile, that features a more rapid flow and a clear asymmetry, due to the overemphasized role of the normal stress differences related to $\lambda_3$ in the modified model.

\begin{figure}[htb]
\centering
\includegraphics[width=0.9\textwidth]{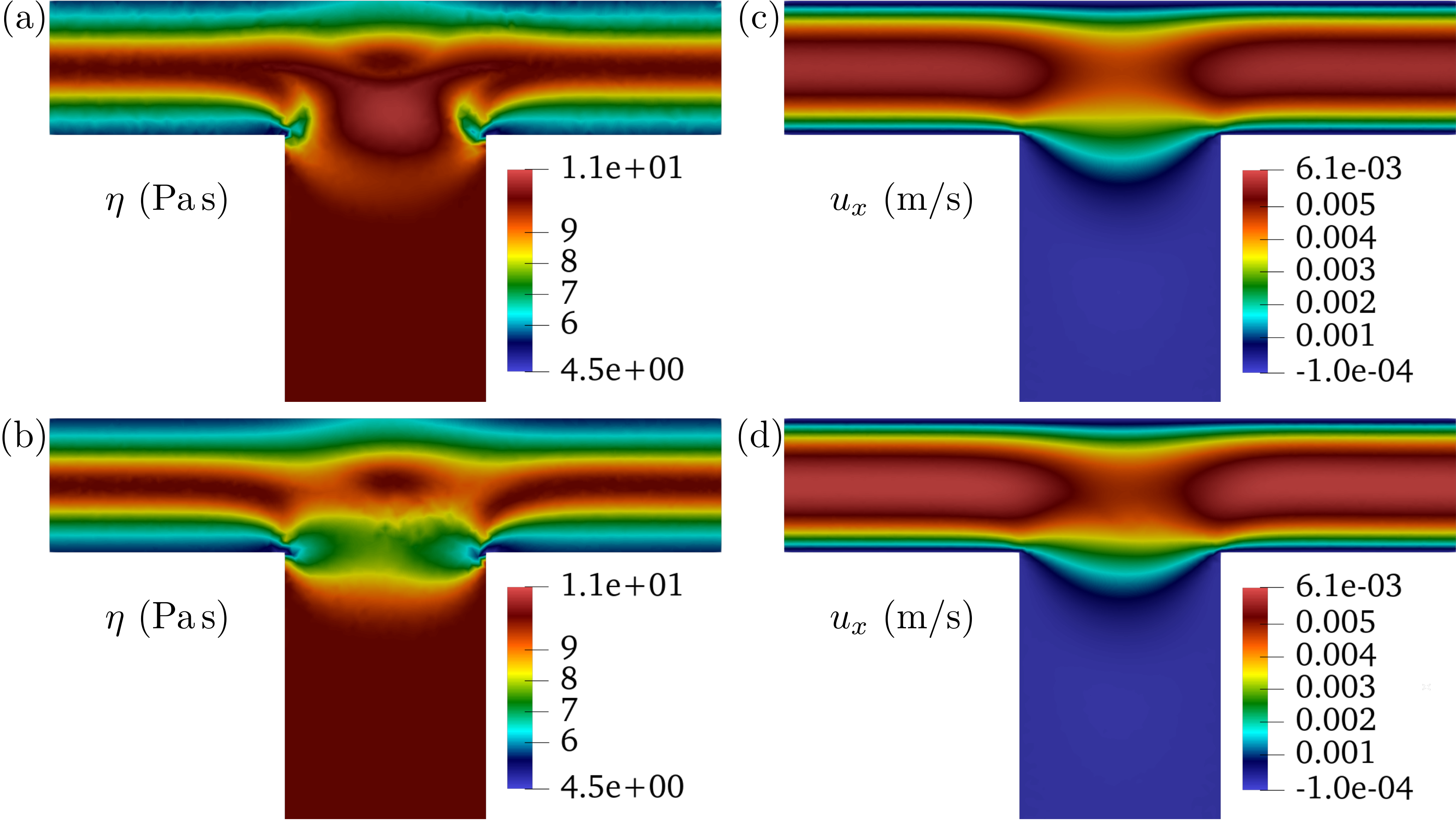}
\caption{\emph{Also in the flow past a deep hole the presence of mixed and extensional flow regions requires a proper treatment of the flow-type dependence of the stress.}
We compare the steady-state viscosity $\eta$ and horizontal velocity $u_x$ computed with (a,c) the full non-Newtonian model and (b,d) the modified one.
The viscosity appears to be significantly different right above the deep hole. The velocity depression is shifted rightward and the flow is globally faster.}
\label{fig:sections_deep}
\end{figure}

\subsection{Flow past a deep hole}

\begin{figure}[t]
    \centering
    \includegraphics[width=0.9\textwidth]{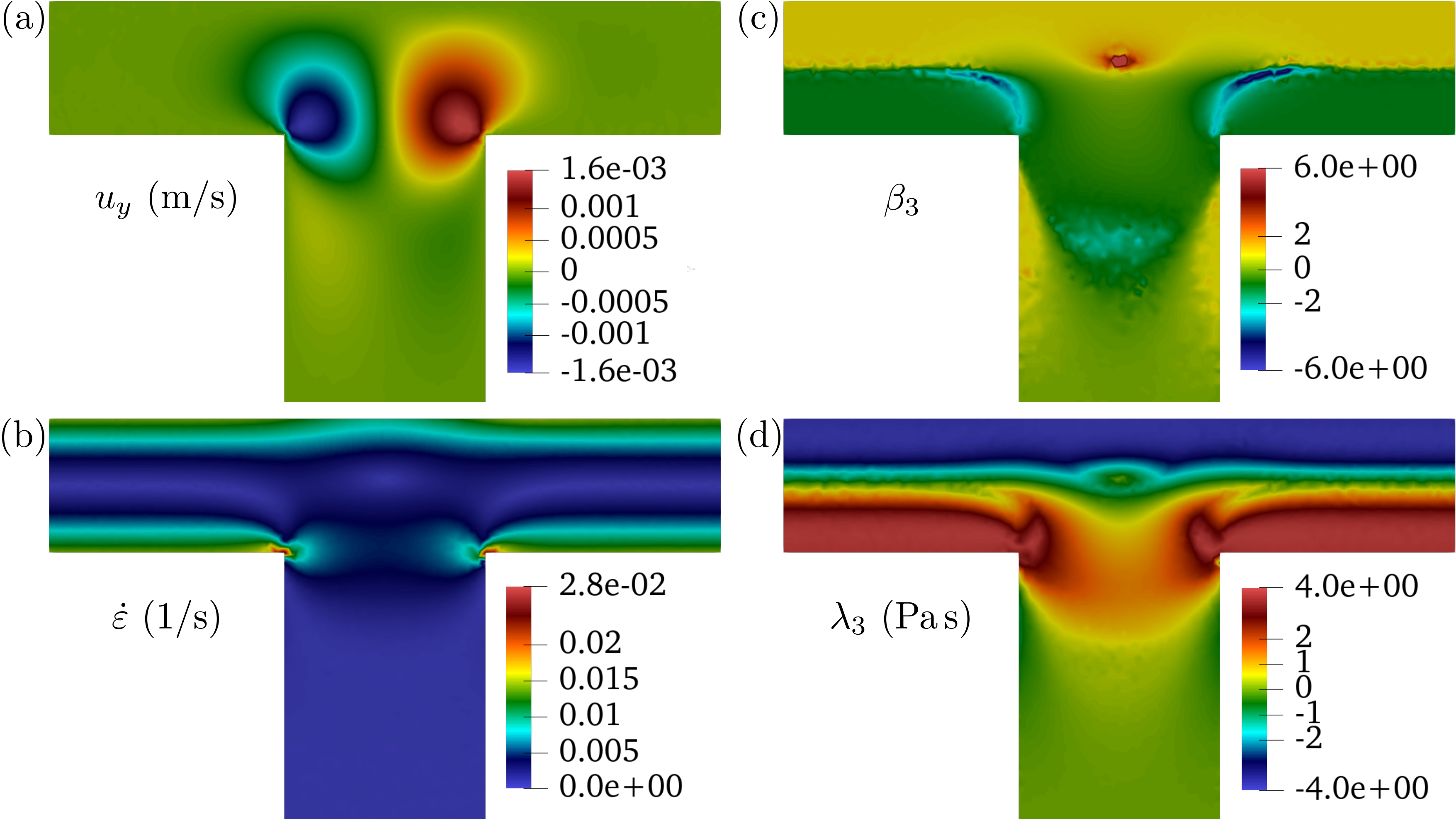}
    \caption{\emph{The flow past a deep hole features a complex distribution of local flow types with a consequent variation of the material response.} 
    (a) From the component $u_y$ of the velocity we can see the region where the deep hole modifies the flow in the upper channel. 
    (b,c) The deformation rate $\ed$ and the flow-type parameter $\beta_3$ highlight the presence of mixed flows and high-vorticity regions. 
    (d) The values of $\lambda_3$ across the domain are a direct consequence of the non-Newtonian rheology and the complexity of the flow. 
    }
    \label{fig:visco_lam_deep}
    \label{fig: streamdeep}
    \label{fig:deep_fields}
\end{figure}

In this section we analyze a second example of flow in a complex geometry.
We consider a channel of length $L= \SI{5}{\meter}$ and width $W= \SI{1}{\meter}$. At a distance of $\SI{1.75}{\meter}$ from the inlet we find the hole of width $W_\mathrm{hole} = \SI{1.5}{\meter}$ and depth $H= \SI{5}{\meter}$.
The discretization employs an unstructured triangular mesh with average diameter of the triangles $h\approx\SI{0.06}{\meter}$.
The time step is $\Delta t= 10^{-3}$ and we let the system evolve to reach a steady state for $T=\SI{0.4}{\second}$ (see Figure~\ref{fig:timeandmesh}, left panel).
We present simulation results for the polymeric case $N_\mathrm{pol}=20$, driving the flow with an outlet-to-inlet pressure gradient $C= \SI{0.3}{\pascal/\meter}$.
The velocity field is again assumed to be periodic, and hence equal at inlet an outlet.

The flow past a deep hole features a complex distribution of local flow types with a consequent variation of the material response.
Also in this case, a comparison of the viscosity and the horizontal velocity obtained with the full and the modified non-Newtonian models (Figure~\ref{fig:sections_deep}) confirms the need for a correct treatment of flow-type-dependent rheologies.
The viscosity appears to be significantly different right above the deep hole. This has a mild but noticeable influence on the flow. 
The velocity depression is shifted rightward and, since the viscosity is lower, the flow is globally faster with the modified model, as seen also from the values of the flow rate reported in Figure~\ref{fig:timeandmesh}.

From the component $u_y$ of the velocity (Figure~\ref{fig:deep_fields}a) we can clearly see the region where the deep hole modifies the flow in the upper channel. Moreover, the presence of a clockwise-rotating vortex can be inferred at the center of the shown portion of the hole. The deformation rate $\ed$ and the flow-type parameter $\beta_3$ highlight the presence of mixed flows and high-vorticity regions (Figure~\ref{fig:deep_fields}b and \ref{fig:deep_fields}c). Specifically, $\beta_3$ confirms the presence of a vortex in the hole. The sharp contrast between yellow and green regions nearby the openings of the upper channel indicates the viscometric nature of the flow far from the hole. The values of $\lambda_3$ across the domain (Figure~\ref{fig:deep_fields}d) are a direct consequence of the non-Newtonian rheology and the complexity of the flow. The same applies to the viscosity shown in Figure~\ref{fig:sections}c.
For this flow, the relevant ranges of dimensionless parameters are $\Rey\in[0.05,0.12]$ and $\varphi\in[-20.8,20.8]$ degrees.
Hence, we have significant non-Newtonian effects while turbulent motions are prevented.

\section{Conclusions and outlook}
\label{Sec6}

We have developed and tested a heterogeneous multi-scale method for the simulation of flows in arbitrary geometries for complex fluids by means of a data-driven reconstruction of the stress tensor.
This is based on employing the results of micro-scale simulations according to a general decomposition of the stress~\cite{giusteri2018theoretical} that allows to correctly align its different components in flows that display a broad spectrum of local flow types.
We have shown that a proper treatment of the flow-type dependence, that was missing in previous multi-scale methods exclusively focused on simple shear rheology, is essential to capture the macroscopic dynamics.

In our case, the micro-scale simulations are non-equilibrium molecular dynamics simulations of polymeric chains with internal FENE bonds and an effective Weeks-Chandler-Andersen interaction potential.
Nevertheless, our coupling scheme does not rely on a specific
simulation strategy neither at the micro-scale nor at the macro-scale.
Indeed, we have chosen to implement the macroscopic continuum simulation with a well-established mixed finite element method based on $\mathsf{P}_2$-$\mathsf{P}_0$ elements for the velocity-pressure pair to deal with the incompressibility constraint and a semi-implicit time integration scheme, but any other computational method to solve for the continuum evolution can in principle be employed. 

Polymeric fluids are well known to display a significant degree of flow-type dependence in their rheological properties. In fact, the different average conformation of the long molecular chains in simple shear and extensional flows 
causes the same material to show opposite trends in the dependence of the viscosity on the rate of deformation.
Our micro-scale simulations reproduce this feature. 
Hence, we were able to assess the relevance of properly taking into account the flow-type dependence of the stress in the macro-scale simulation.
From the results obtained with our multi-scale approach in paradigmatic geometries, it is clear that the predictions of a model that suppresses the flow-type dependence are not reliable.

We performed micro-scale simulations imposing simple shear and planar extension as average kinematic conditions under which we evaluate the stress response. From these, we constructed the rheological functions in mixed flows by means of a simple interpolation and extrapolation procedure.
In view of the impact on macroscopic flows in non-viscometric geometries of the local flow conditions, we plan to expand the micro-scale simulations to obtain data under mixed and rotational flow conditions.
This will require the implementation of a suitable generalization of the Kraynik--Reinelt boundary conditions.

In fluids composed by molecules with longer microscopic relaxation times, it may happen that the local flow type experienced by a fluid element changes fast along streamlines.
The method that we presented can in principle be extended to take this into account by devising MD simulations for which the flow type of the imposed background motion is slowly varying. The macroscopic material functions would then depend also on the derivative along streamlines of the flow-type parameter.
Other directions for future work are the testing of our method in parameter ranges where elastic instabilities may be observed, that may lead to the presence of highly oscillatory material coefficients in space and time, and the implementation of an on-the-fly coupling between macro-scale and micro-scale simulations, to be able to sample the space of kinematic conditions in a problem-driven fashion.

\section*{Acknowledgments}
G.G.G.\ and F.T.\ acknowledge the support of the Italian National Group of Mathematical Physics (GNFM-INdAM) through the funding scheme \emph{GNFM Young Researchers' Projects 2020}.
The research of M.L.-M.\ and L.Y.\ was  supported by the German Science Foundation (DFG) within the Project number 233630050-TRR 146, C5 project. They also gratefully acknowledge
the computing time granted on the HPC cluster
Mogon at Johannes Gutenberg University Mainz. M.L.-M.\ is grateful to the Gutenberg
Research College of the Johannes Gutenberg University Mainz for the research support.

\appendix

\section{Fitting procedure for the micro-scale NEMD data}
\label{appendice}

We report here the details about the fitting of micro-scale data that we performed using Gaussian Process Regression (GPR). Since some choices are involved in the procedure, it seems appropriate to give a brief account of ours.
GPR is a Bayesian approach to regression based on machine learning techniques.
The fitting is done through a Gaussian stochastic process  $f(x; \mathbf{w})$ that evolves along the variable $x$ and depends on a list of hyper-parameters $\mathbf{w}$.
The observed data are a finite number $N$ of pairs $(x_i,y_i)_{i=1}^N$.
The functional form of the Gaussian process is specified by its mean and covariance functions, $m({x}; {\bf w}) = \mathbb{E}[f({x}; {\bf w})]$ and $K({x}, { x'}; {\bf w}) = \mathbb{E}\big[\big(f({x};{\bf w})-m({x};{\bf w})\big)\big(f({x'};{\bf w})-m({x'};{\bf w})\big)\big]$,
where $\mathbb E$ denotes the expected value.
A fundamental feature of GPR is that it does not produce a definite value of the fitting parameters $\mathbf{w}$, but considers them random variables and seeks to determine suitable mean and variance for their distribution.
This is achieved by an iterative optimization that takes into account the likelihood of the parameters distributions given the knowledge of the observed data~\cite{williams1996gaussian}.

What one needs to specify is the type of mean and covariance functions to be used for the GPR. Typical choices for the former are polynomial functions of $x$ with coefficients given by $\mathbf{w}$, while for the latter are Radial Basis Functions
\begin{equation*}
    K_\mathrm{RBF}({x},{x'};\sigma, l) \equiv \sigma^2 \exp{\left(-\frac{|{x}-{x'}|^2}{2l^2}\right)}
\end{equation*}
with a set of two hyper-parameters ${\bf w}=(\sigma, l)$, or 
the Matérn covariance of degree $\nu$ given by
\begin{equation*}
    K_{\mathrm{M},\nu}({x},{x'};\sigma,\varrho)\equiv \sigma^2 \frac{2^{1-\nu}}{\Gamma(\nu)}\left(\sqrt{2\nu} \frac{|{x}- {x'}|}{\varrho}\right)^\nu K_\nu \left(\sqrt{2\nu} \frac{|{x}- {x'}|}{\varrho}\right),
\end{equation*}
where $\Gamma$ is the Euler gamma function and $K_\nu$ is the modified Bessel function of the second kind.
In our treatement, we took as $x$ variable the quantity $\ed$ or $\log_{10}(\ed)$, as $y$ variable we took $\eta$, $\log_{10}(\eta)$, or $\lambda_3$. We tested regressions with mean function zero, 
constant, or polynomial up to degree four and, for the covariance function, RBF or Mat\'ern kernel 
of degree up to 5 and picked the combinations that give the best results after optimization.

\section{Multi-scale coupling in the code}
\label{app:code}

The purpose of this section is to present the portion of code that implements the multi-scale coupling in the context of an otherwise standard Finite Element simulation of the flow equations (see~\cite{langtangen2016solving} for an introduction to the library FEniCS). 
The peculiarity of our method concerns the integration of NEMD data with the continuum solver and the construction of the stress tensor based on the general representation~\eqref{stressreco}.

In our code, \verb|numpy| arrays contain the interpolated curves $\eta_{\mathrm{sh}}$, $\lambda_{3,\mathrm{sh}}$ and $\eta_{\mathrm{ext}}$, while $\lambda_{3,\mathrm{ext}}$ is simply zero.
In particular, the first column of the array \verb|S| contains the values of $\ed$ on which the Carreau function or GPR is sampled and the second column the corresponding expected values of $\eta_\mathrm{sh}$.
For an efficient use of those data in the stress computation, we transform them in projections on a Finite Element space defined on a one-dimensional mesh, the nodes of which are given by the sampled values of $\ed$.
Considering, for instance, the curve $\eta_\mathrm{sh}$, it becomes the function \verb|etas| in the space \verb|E| of $\mathsf{P}_1$ elements defined on the mesh \verb|mesh_etas|. Its construction is performed with the following lines of code. Similar commands are used to define the functions \verb|etae| and \verb|lambda3s| that represent $\eta_\mathrm{ext}$ and $\lambda_{3,\mathrm{sh}}$.

\begin{lstlisting}[language=Python]
# 1D mesh for eta_sh

import numpy as np
from mshr import *

nodes = S[:,0]
n_nodes = nodes.shape[0]
n_cells = n_nodes - 1
cells = np.arange(n_cells)

# The geometry is inherited by a standard mesh
mesh0 = UnitIntervalMesh(n_cells)
gdim = mesh0.geometry().dim()
tdim = mesh0.topology().dim()
c_type = mesh0.type()
c_str = CellType.type2string(c_type.cell_type())

# Mesh nodes are specified from data
mesh_etas = Mesh()
editor = MeshEditor()
editor.open(mesh_etas, c_str, tdim, gdim)
editor.init_vertices(n_nodes)
editor.init_cells(n_cells)
[editor.add_vertex(int(i),[n]) for i,n in enumerate(nodes)]
[editor.add_cell(i,[i,i+1]) for i in range(cells.shape[0])]
editor.close()

# Definition of FE space and function
E = FunctionSpace(mesh_etas, 'P', 1)
etas= Function(E)
etas.vector().set_local(np.flip(S[:,1], axis=0))  
etas.set_allow_extrapolation(True)
\end{lstlisting}

The second portion of code relevant to the multi-scale coupling is the local computation of the kinematic parameters $\ed$ and $\beta_3$ and the reconstruction, at each time step, of the spatial profile of the stress coefficients $\eta$ and $\lambda_3$ by means of the material functions $\tilde{\eta}(\ed,\beta_3)$ and $\tilde{\lambda}_3(\ed,\beta_3)$.
This is accomplished through the following definitions, where \verb|mesh| is the two-dimensional triangular mesh defined on the flow domain, and \verb|u| is a three-component variable containing the velocity and pressure fields.

\begin{lstlisting}[language=Python]
from fenics import *

def Max(a, b):
    return (a+b+abs(a-b))/Constant(2)
    
def Min(a, b):
    return (a+b-abs(a-b))/Constant(2)

# Epsilondot from the velocity u
def eps(u):
    return sqrt(0.5*((u[0].dx(0))**2+(u[1].dx(1))**2)+0.25*(u[1].dx(0)+u[0].dx(1))**2)

# Epsilondot truncated for viscosity in shear
def eps_es(u):       
    a = nodes[0]
    b = nodes[-1]
    return Max(a,Min(b, eps(u)))

# Epsilondot truncated for viscosity in extension
def eps_ee(u):
    a = nodes_e[0]
    b = nodes_e[-1]
    return Max(a,Min(b, eps(u)))

# Epsilondot truncated for lambda_3 in shear
def eps_ls(u):
    a = nodes_l[0]
    b = nodes_l[-1]
    return Max(a,Min(b, eps(u)))

# Beta3 from the velocity u truncated between -1 and 1
def beta(u):
    ux = u[0]
    uy = u[1]
    uxy= ux.dx(1)
    uyx= uy.dx(0)
    rot= uyx - uxy
    return Max(-1, Min(1, rot/(2*Max(1.e-12, eps(u)))))
    
# Initialize functions
Q = FunctionSpace(mesh, 'P', 1)
epsilondot_ = Function(Q)
beta3_ = Function(Q)
eta = Function(Q)
lambda3 = Function(Q)
\end{lstlisting}

At each time step, we need to compute the new stress coefficients based on the velocity field \verb|u_n| obtained in the previous step. These will be used to update the stress. To this end, we apply the interpolation between simple shear and extensional data according to \eqref{eq:eta}--\eqref{eq:lambda}.

\begin{lstlisting}[language=Python]
# Within the time evolution loop
    # Update kinematic parameters
    edot_es= project(eps_es(u_n), Q).vector().get_local()
    edot_ee= project(eps_ee(u_n), Q).vector().get_local()
    edot_ls= project(eps_ls(u_n), Q).vector().get_local()
    beta3= project(beta(u_n), Q).vector().get_local()
    # Evaluate material functions    
    etas_temp= np.zeros(edot_es.shape[0])
    etae_temp= np.zeros(edot_ee.shape[0])
    eta_temp= np.zeros(edot_es.shape[0])
    lambda3_temp= np.zeros(edot_ls.shape[0])
    for i in range(edot_es.shape[0]):
        etas_temp[i]= etas(edot_es[i])
        etae_temp[i]= etae(edot_ee[i])
        lambda3_temp[i]= lambda3s(edot_ls[i])
    eta_temp= np.absolute(beta3)*etas_temp + (1-np.absolute(beta3))*etae_temp
    lambda3_temp= beta3*lambda3_temp
    # Casting the data into functions on the 2D mesh
    eta.vector().set_local(eta_temp)
    lambda3.vector().set_local(lambda3_temp)
\end{lstlisting}

\end{document}